\documentclass[nonacm,sigplan, screen,x11names]{acmart}

\usepackage{amsmath}
\usepackage{parskip}
\usepackage{xspace}
\usepackage{hyperref}
\usepackage{cleveref}
\usepackage{csquotes}

\usepackage{mdframed}
\usepackage[]{xcolor}
\usepackage{todonotes}
\usepackage{enumitem}
\usepackage{multicol}
\usepackage{multirow}

\usepackage{wasysym}

\usepackage{tikz}
\usetikzlibrary{arrows,fit,decorations.pathreplacing,calligraphy}
\usepackage{xcolor}
\definecolor{mygreen}{HTML}{3C8031}
\definecolor{myorange}{HTML}{F58137}

\tikzset{
	contrapos/.style={
		draw=mygreen,
		very thick,
		dotted
	},
	galois/.style={
		draw=myorange,
		very thick
	},
	implic/.style={
		draw=myorange,
		very thick
	},
}

\title[Hoare-Like Triples and Kleene Algebras with Top and Tests]{Hoare-Like Triples and\\ Kleene Algebras with Top and Tests}
\subtitle{Towards a Holistic Perspective on Hoare Logic, Incorrectness Logic, and Beyond}

\author{Lena Verscht}
\affiliation{%
	\institution{Saarland University}
	\country{}
}
\email{lverscht@cs.uni-saarland.de}

\author{Benjamin Kaminski}
\affiliation{%
	\institution{\hspace{1.2em}Saarland University{\hspace{0.8em}}and{\hspace{0.8em}}University College London}
	\country{}
}
\email{kaminski@cs.uni-saarland.de}

\begin{document}

\newcommand{\hoare}[3]{\left\{ #1 \right\} \mathrel{#2} \left\{ #3 \right\}}
\newcommand{\incorrectness}[3]{\left[ #1 \right] \mathrel{#2} \left[ #3 \right]}


\newenvironment{narrowbluebox}{%
	\medskip%
	\begin{mdframed}[backgroundcolor=DodgerBlue3!10!white, linecolor=gray, linewidth=.25pt, skipabove=0pt, skipbelow=1.5\topsep, innertopmargin=1em, innerbottommargin=1em, innerleftmargin=1em, innerrightmargin=1em]%
}{%
	\end{mdframed}
}

\newcommand{\morespace}[1]{~{#1}~}
\newcommand{\qmorespace}[1]{\quad{#1}\quad}
\newcommand{\qqmorespace}[1]{\qquad{#1}\qquad}

\newcommand{\qiff}{\qmorespace{\textnormal{iff}}}
\newcommand{\qqiff}{\qqmorespace{\textnormal{iff}}}

\newcommand{\todoil}[1]{\todo[inline]{#1}}

\newcommand{\pre}{\ensuremath{b}}
\newcommand{\post}{\ensuremath{c}}
\newcommand{\program}{\ensuremath{p}}

\newcommand{\proglogin}{\ensuremath{\program_{\text{login}}}}

\newcommand{\KAT}{\textup{\textsf{KAT}}\xspace}
\newcommand{\CompKAT}{\textup{\textsf{CompKAT}}\xspace}
\newcommand{\TopKAT}{\textup{\textsf{TopKAT}}\xspace}

\newcommand{\wppname}{\ensuremath{\mathsf{wp}}}
\newcommand{\wlpname}{\ensuremath{\mathsf{wlp}}}

\newcommand{\wpp}[2]{\ensuremath{\mathsf{wp}\llbracket #1 \rrbracket\left( #2 \right)}}
\newcommand{\wlp}[2]{\ensuremath{\mathsf{wlp}\llbracket #1 \rrbracket\left( #2 \right)}}

\newcommand{\awppname}{\ensuremath{\mathsf{awp}}\xspace}
\newcommand{\awlpname}{\ensuremath{\mathsf{awlp}}\xspace}
\newcommand{\dwppname}{\ensuremath{\mathsf{dwp}}\xspace}
\newcommand{\dwlpname}{\ensuremath{\mathsf{dwlp}}\xspace}

\newcommand{\awpp}[2]{\ensuremath{\mathsf{awp}\llbracket #1 \rrbracket\left( #2 \right)}}
\newcommand{\awlp}[2]{\ensuremath{\mathsf{awlp}\llbracket #1 \rrbracket\left( #2 \right)}}
\newcommand{\dwpp}[2]{\ensuremath{\mathsf{dwp}\llbracket #1 \rrbracket\left( #2 \right)}}
\newcommand{\dwlp}[2]{\ensuremath{\mathsf{dwlp}\llbracket #1 \rrbracket\left( #2 \right)}}

\newcommand{\sppname}{\ensuremath{\mathsf{sp}}}
\newcommand{\slpname}{\ensuremath{\mathsf{slp}}}

\newcommand{\spp}[2]{\ensuremath{\mathsf{sp}\llbracket #1 \rrbracket\left( #2 \right)}}
\newcommand{\slp}[2]{\ensuremath{\mathsf{slp}\llbracket #1 \rrbracket\left( #2 \right)}}

\newcommand{\asppname}{\ensuremath{\mathsf{asp}}\xspace}
\newcommand{\aslpname}{\ensuremath{\mathsf{aslp}}\xspace}
\newcommand{\dsppname}{\ensuremath{\mathsf{dsp}}\xspace}
\newcommand{\dslpname}{\ensuremath{\mathsf{dslp}}\xspace}

\newcommand{\aspp}[2]{\ensuremath{\mathsf{asp}\llbracket #1 \rrbracket\left( #2 \right)}}
\newcommand{\aslp}[2]{\ensuremath{\mathsf{aslp}\llbracket #1 \rrbracket\left( #2 \right)}}
\newcommand{\dspp}[2]{\ensuremath{\mathsf{dsp}\llbracket #1 \rrbracket\left( #2 \right)}}
\newcommand{\dslp}[2]{\ensuremath{\mathsf{dslp}\llbracket #1 \rrbracket\left( #2 \right)}}

\maketitle

\section{Introduction}
\begin{flushright}
	\enquote{Program correctness and incorrectness are\\ two sides of the same coin.}\\[.5em] 
	--- \citet{o2019incorrectness}
\end{flushright}

We argue that the object of discourse is not a coin and that it has at least three sides or rather \emph{dimensions}:%
\begin{enumerate}
	\item
		\emph{correctness} vs.~\emph{incorrectness}
	\item
		\emph{totality} vs.~\emph{partiality}
	\item
		\emph{reachability} vs.~\emph{unreachability}
\end{enumerate}%
We will explore how one can pigeonhole \emph{total} and \emph{partial correctness}, as well as \emph{incorrectness} into this view and explore further program properties that emerge from exhausting all combinations of the above dimensions.
We will furthermore explore how to express validity of those properties in the language of Kleene algebras with top and tests~\cite{zhang2022incorrectness}.

\section{Canonical Exegeses of Triples}

The central notion of Hoare logic~\cite{hoare1969axiomatic} are triples of the form $\hoare{\pre}{\program}{\post}$, where $\program$ is a program, $\pre$ is a \emph{precondition} on initial states, and $\post$ is a \emph{postcondition} on final states.
To give meaning to a triple, it must be \emph{exegeted} when a triple is considered \emph{valid} and when it is not.
For Hoare logic, there are two such exegeses: validity for \emph{total} and for \emph{partial} correctness.

\paragraph{\textsc{\textup{Exegesis I:}} Partial Correctness.} 
$\hoare{\pre}{\program}{\post}$ is valid for \emph{partial correctness} iff all executions of $\program$ that start in~$\pre$ can only ever terminate in~$\post$.
The execution \emph{is} allowed to diverge.
For example, validity for partial correctness of%
\begin{align*}
	\hoare{\text{wrong password}}{\quad\proglogin\quad}{\text{login failed}}
\end{align*}%
specifies that when the user enters a wrong password, $\proglogin$ will either diverge or terminate with a failed login.
In particular, successful login is impossible with the wrong password.

\paragraph{\textsc{\textup{Exegesis II:}} Total Correctness.} 
$\hoare{\pre}{\program}{\post}$ is valid for \emph{total correctness} iff all executions of $\program$ that start in $\pre$ definitely \emph{terminate} and they do so in~$\post$.
Reusing the example above, validity for total correctness of%
\begin{align*}
	\hoare{\text{correct password}}{\quad \proglogin \quad}{\text{login successful}}~,
\end{align*}%
specifies that when the user enters the correct password, the login will definitely be successful.

It is noteworthy that both partial and total correctness are about \emph{coreachability of all initial states}:
$\hoare{\pre}{\program}{\post}$ is valid if \emph{all} initial states in $\pre$ are (partially) mapped by $\program$ \emph{into} (but not necessarily \emph{onto}) the final states in $\post$.

\paragraph{Weakest (Liberal) Preconditions.}

Proposed by Dijkstra~\cite{dijkstra1976discipline}, \emph{weakest (liberal) preconditions} are a very versatile alternative to Hoare logic and led to numerous extensions \cite{morgan1996probabilistic,kaminski2018weakest,kaminski2019advanced,zhang2022quantitative}.
Given only a program $\program$ and a postcondition $\post$, the  \emph{weakest liberal precondition} is the weakest predicate $\wlp{\program}{\post}$, such that all executions of $\program$ that start in $\wlp{\program}{\post}$ either diverge or terminate in $\post$.
One can (re)define partial correctness in terms of weakest liberal preconditions:
$\hoare{\pre}{\program}{\post}$ is valid for partial correctness iff $\pre \implies \wlp{\program}{\post}$.

For \emph{total} correctness, the weakest precondition is the weakest predicate~$\wpp{\program}{\post}$, such that all executions of $\program$ that start in $\wpp{\program}{\post}$ definitely terminate and they do so in $\post$.
One can define total correctness in terms of weakest preconditions:
$\hoare{\pre}{\program}{\post}$ is valid for total correctness iff $\pre \implies \wpp{\program}{\post}$.%

\paragraph{\textsc{\textup{Exegesis III:}} Incorrectness.} 

Incorrectness triples were introduced by~\citet{de2011reverse} and later independently reintroduced by~\citet{o2019incorrectness}.
$\incorrectness{\pre}{\program}{\post}$ is valid for incorrectness iff all executions of $\program$ that terminate in $\post$ could have started in $\pre$.
In other words: iff $\program$ maps (a subset of) $\pre$ \emph{onto} $\post$.
In that sense, incorrectness is really about \emph{reachability of all final states}.
If $\post$ is a set of error states, then validity of $\incorrectness{\pre}{\program}{\post}$ for incorrectness indeed proves that executing $\program$ on $\pre$ is not safe, because doing so can reach an error in~$\post$.
For example, validity for incorrectness of%
\begin{align*}
	\incorrectness{\text{correct password}}{\quad \proglogin \quad}{\text{program crash}}
\end{align*}%
specifies that $\proglogin$ can crash on entering a correct password.

\paragraph{Strongest Postconditions.}

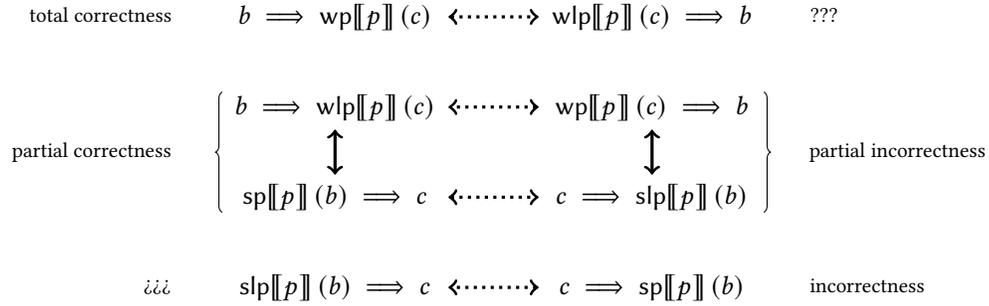
\begin{figure*}[t!]
	\begin{tikzpicture}[mynode/.style={minimum width=3cm},node distance=.6cm and 1.2cm]
		\draw[help lines, draw=none,use as bounding box] (-5.5, 0.2) grid (9.7,-3.8);
		
		\node[mynode](awpLB) {$\pre \implies \wpp{\program}{\post}$};
		\node[mynode, left= of awpLB,xshift=1.1cm] (total corr) {\footnotesize total correctness\vphantom{g}};
		
		\node[mynode,right=of awpLB]  (dwlpUB) {$\wlp{\program}{\post} \implies \pre$};
		\node[mynode, right=of dwlpUB,xshift=-1.925cm] (qqq) {\footnotesize ???\vphantom{g}};
		\node[mynode,below=of awpLB] (dwlpLB) {$\pre \implies \wlp{\program}{\post}$};
		\node[mynode,right=of dwlpLB] (awpUB) {$\wpp{\program}{\post} \implies \pre$};
		
		\node[mynode,below=of dwlpLB] (aspUB) {$\spp{\program}{\pre} \implies \post$};
		\node[mynode,below=of aspUB]  (dslpUB) {$\slp{\program}{\pre} \implies \post$};
		\node[mynode, left= of dslpUB,xshift=1.85cm] (uququq) {\footnotesize ?`?`?`\vphantom{g}};
		
		\draw [decorate, decoration = {brace}] (-1.5,-2.6) -- node[pos=0.5] (middletoken1) {} (-1.5,-1);
		\node[mynode, left=of middletoken1,xshift=1.1cm] (part corr) {\footnotesize partial correctness\vphantom{g}};
		
		\node[mynode,right=of aspUB] (dslpLB) {$\post \implies \slp{\program}{\pre}$};
		
		\node[mynode,below=of dslpLB] (aspLB) {$\post \implies \spp{\program}{\pre}$};
		\node[mynode, right=of aspLB,xshift=-1.35cm] (total incorr) {\footnotesize incorrectness\vphantom{g}};
		
		\draw [decorate, decoration = {brace}] (5.7,-1) -- node[pos=0.5] (middletoken2) {} (5.7,-2.6);
		\node[mynode, right=of middletoken2,xshift=-1.05cm] (part corr) {\footnotesize partial incorrectness\vphantom{g}};
		
		\draw[contrapos] (awpUB) edge[<->] (dwlpLB);
		\draw[contrapos] (awpLB) edge[<->] (dwlpUB);
		\draw[contrapos] (aspUB) edge[<->] (dslpLB);
		\draw[contrapos] (aspLB) edge[<->] (dslpUB);
		
		\draw[galois] (aspUB) edge[<->] (dwlpLB);
		\draw[galois] (awpUB) edge[<->] (dslpLB);
	\end{tikzpicture}
	\caption{
		An overview of the exegeses of Hoare triples and how they are related.
		The dotted green arrows in the middle symbolize contrapositives and the orange arrows symbolize equivalences through Galois connections.
	}
	\label{fig:landscape}
\end{figure*}%
Given only a program $\program$ and a \emph{pre}condition $\pre$, the \emph{strongest postcondition}~\cite{DBLP:books/daglib/0067387} is the strongest predicate $\spp{\program}{\pre}$, such that any state in $\spp{\program}{\pre}$ is reachable from $\pre$ by executing $\program$.
We can redefine incorrectness in terms of strongest postconditions~\cite{zhang2022quantitative}:
\mbox{$\incorrectness{\pre}{\program}{\post}$} is valid for incorrectness iff \mbox{$\post \implies \spp{\program}{\pre}$}.

$\sppname$ has a Galois connection to $\wlpname$, namely%
\begin{align*}
	\pre \implies \wlp{\program}{\post} \qqiff \spp{\program}{\pre} \implies \post~. \tag{$\dagger$}
\end{align*}
Because of this, we can also define partial correctness in terms of strongest postconditions: 
$\hoare{\pre}{\program}{\post}$ is valid for partial correctness iff $\spp{\program}{\pre} \implies \post$.
Notice that the difference between incorrectness and partial correctness is merely the direction of the implication between $\post$ and $\spp{\program}{\pre}$.

\section{Completive Exegeses of Triples}

Given a triple of precondition~$\pre$, program~$\program$, and postcondition~$\post$, we have so far seen three exegesis of such triples, all given in terms of $\wlpname$/$\wppname$/$\sppname$ and implications:%
\begin{center}%
	\renewcommand{\arraystretch}{1.25}
	\begin{tabular}{r@{\qquad}l}
		total correctness				& $\pre \implies \wpp{\program}{\post}$\\[.5em]
		\multirow[c]{2}{*}{partial correctness}	& $\pre \implies \wlp{\program}{\post}$\\
									& $\spp{\program}{\pre} \implies \post$\\[.5em]
		incorrectness					& $\post \implies \spp{\program}{\pre}$
	\end{tabular}%
\end{center}%
From inspecting this table, we can make out three degrees of freedom one has in exegeting triples:
\begin{enumerate}
	\item 
		The difference between total correctness and partial correctness is the usage of $\wppname$ vs.~$\wlpname$.
	\item
		The difference between partial correctness and incorrectness is the direction of the implication.
	\item
		The difference between total correctness and incorrectness is $\wppname$ vs.~$\sppname$.
\end{enumerate}%
This immediately gives rise to two questions: 
What about $\wpp{\program}{\post} {\implies} \pre$ and $\wlp{\program}{\post} {\implies} \pre$? 
And can we define strongest liberal postconditions?

\paragraph{Strongest Liberal Postconditions.}

Indeed, these can be sensibly defined~\cite{zhang2022quantitative}.
The difference between $\wlpname$ and $\wppname$ is that the liberal variant additionally contains all states that diverge, i.e.\ those \emph{initial states for which there exists no final state} in which the execution of $\program$ could terminate.
Dually, $\slp{\program}{\pre}$ contains also all states that \emph{cannot be reached at all} by executing $\program$, i.e.\ those \emph{final states for which there exists no initial state} in which the execution of $\program$ could have started.

$\slpname$ also has a Galois connection~\cite{zhang2022quantitative}, but to $\wppname$, namely%
\begin{align*}
	\wpp{\program}{\post} \implies \pre \qqiff \post \implies \slp{\program}{\pre}~.\tag{$\ddagger$}
\end{align*}
Exhaustively combining $\wppname$, $\wlpname$, $\sppname$, and $\slpname$ with all implication directions gives a completed canon of in total 8 different exegeses of triples, four of which reduce to only two via Galois connections, see \Cref{fig:landscape}.%

\paragraph{\textsc{\textup{Exegesis IV:}} Partial Incorrectness.} 

The difference between total and partial correctness is the use of $\wppname$ vs.~$\wlpname$.
Consequently, when we replace $\sppname$ by $\slpname$ in the defining implication $\post \implies \spp{\program}{\pre}$ of incorrectness, we should arrive at the exegesis of \emph{partial incorrectness}, as suggested (but not further explored) by \citet{zhang2022quantitative}:
$[\pre] \program [\post]$  is valid for \emph{partial incorrectness} iff $\post \implies \slp{\program}{\pre}$ (and by~($\ddagger$) equivalently iff $\wpp{\program}{\post} \implies \pre$), which means that all states in $\post$ are either unreachable (from anywhere) or reachable from $\pre$ by executing $\program$.
For example, validity for partial incorrectness of
\begin{align*}
	\incorrectness{\text{wrong password}}{\quad \proglogin \quad}{\text{login successful}}
\end{align*}
specifies that any state where the login was successful is either entirely unreachable, or reachable by entering a wrong password, thus constituting a potential bug.

\paragraph{\textsc{\textup{Exegesis V and VI}}} 

There are two remaining triple exegeses, defined by $\wlp{\program}{\post} \implies \pre$ and $\slp{\program}{\pre} \implies \post$, respectively.
These have also been suggested by \citet{zhang2022quantitative}, but not further explored.

\paragraph{Contrapositives}

Besides some actual \emph{equivalences} via Galois connections, some triples are contrapositives of each other~\cite{zhang2022quantitative}.
In fact, the entire left-hand-side (in some sense the \emph{correctness side}) is the contrapositive of the right-hand-side (the \emph{incorrectness side}) in \Cref{fig:landscape}.
For example,%
\begin{align*}
	\pre \implies \wlp{\program}{\post} \quad \text{iff} \quad \wpp{\program}{\neg \post} \implies \neg \pre
\end{align*}%
and hence partial correctness is the contrapositive of partial incorrectness and so both describe in some sense the same class of specifications.
We would claim, however, that it might be more intuitive for a working programmer to specify either a partial incorrectness triple or a partial correctness triple, depending on the property they have in mind.

\section{Kleene Algebra with Top and Tests}

\emph{Kleene algebra with tests} (\KAT), introduced by \citet{kozen1997kleene}, is another alternative for specifying and reasoning about program properties.
\KAT terms are generalized regular expressions over a \emph{two-sorted alphabet}  comprising (i)~programs ($p,q,\ldots$) and (ii)~tests ($b,c,\ldots$).
We interpret these symbols as relations: a program $\program$ relates (maps) initial states to final states by its execution.
Initial states on which~$\program$ diverges are not related with any final states.
Similarly, unreachable final states are not related with any initial states.
A test $\pre$ maps initial states satisfying $\pre$ to themselves, and those not satisfying $\pre$ to no state.
In this sense, tests act as filters.

Testing for $\textsf{false}$ is denoted by $0$ and gives the \emph{empty} relation.
In the lattice of relations (ordered by set inclusion), $0$ is the least/bottom element.
In \emph{Kleene algebra with top and tests}~(\TopKAT), we now additionally add a $\top$ element to \KAT, which codes for the \emph{universal} relation relating every initial state to every final state~\cite{zhang2022incorrectness}.

Composing symbols means composing relations:
For example, the term $bpqc$ intuitively means: first test for $\pre$, then execute~$p$ (but only on states that satisfy $\pre$), then execute~$q$, then test for $\post$.
Executions that fail to satisfy initially~$\pre$ or finally $\post$ are filtered out and not part of the resulting relation.

\paragraph{Partial Correctness in \TopKAT}
We can express partial correctness by the \TopKAT equation \mbox{$\top bpc = \top bp$}.
Prepending the top element on both relations in the equation amounts to comparing their \emph{codomains}, i.e.\ their \emph{final states}.
Hence, $\top bpc = \top bp$ expresses that the set of final states in $\post$ in which $\program$ can terminate starting from $\pre$ is exactly those final states in which $\program$ can terminate from $\pre$ at all.
No statement is made about initial states in $\pre$ on which $\program$ diverges.

\paragraph{Incorrectness in \TopKAT}
It is not possible to express incorrectness in \KAT, but this \emph{is} possible in \TopKAT \cite{zhang2022incorrectness}, namely by $\top bpc = \top c$.
On the left-hand-side of the equation, we select all final states in $\post$ that were reachable by executing~$\program$ on~$\pre$.
On the right-hand-side, we select \emph{all} final states in~$\post$.

\paragraph{Nondeterminism and Correctness in \KAT}
So far, we have kept quiet about nondeterminism, but in Kleene algebras it is very natural to model nondeterministic programs.
Considering total correctness of $\hoare{\pre}{\program}{\post}$, one then has to decide: 
(i) must \emph{all} execution paths emerging from any state in~$\pre$ terminate in~$\post$, or (ii) must there merely exist \emph{some} path that does so?
The former is called the \emph{demonic} exegesis of nondeterminism while the latter is called \emph{angelic}.
On that note, \citet{zhang2022quantitative} always interpret nondeterminism \emph{angelically} for their \emph{non}liberal pre- and postconditions, and \emph{demonically} for their liberal ones.
Only this way, they obtain the Galois connections ($\dagger$) and ($\ddagger$).
However, the standard notion of total correctness is \emph{demonic} and not \emph{angelic}.

Unfortunately, it is known that \emph{demonic} total correctness is inexpressible in \KAT\cite{von2002kleene} because it lacks a way of reasoning about nontermination: 
Nonterminating executions are just not part of the resulting relation.
For the angelic variant, however, $bpc\top = b\top$ indeed expresses that from all intial states in $\pre$, it is \emph{possible} for $\program$ to terminate in $\post$, thus:%
\vspace*{-.25em}%
\begin{narrowbluebox}%
\abovedisplayskip=-.25em%
\begin{align*}
	\begin{array}{c}
		\hoare{\pre}{\program}{\post} \textnormal{ is valid for angelic total correctness}\\[.5em]
		\textnormal{iff}\\[.5em]
		bpc\top = b\top
	\end{array}
\end{align*}%
\normalsize%
\end{narrowbluebox}%
\vspace*{-.25em}%
This is, to the best of our knowledge, a novel result.
Moreover, this result renders \emph{all} triples depicted in \Cref{fig:landscape} expressible in \TopKAT in a syntactically very similar manner:
\begin{center}%
	\renewcommand{\arraystretch}{1.25}
	\begin{tabular}{r@{\qquad}l}
		total correctness	& $bpc\top = b\top$\\[.25em]
		partial correctness	& $\top bpc = \top bp$\\[.25em]
		incorrectness		& $\top bpc = \top c$
	\end{tabular}%
\end{center}%
The opposite-hand-sides are expressible via contrapositives.
For example, validity of $\incorrectness{\pre}{\program}{\post}$ for partial incorrectness is expressible via $bpc \top= pc \top$.
And more equations fit into this pattern, for example $\top bpc = \top pc$  and $bpc \top= bp\top $.
Interestingly, these two express the angelic variants of partial correctness and incorrectness, respectively, and thus lead to further exegeses of triples.

\section{Ongoing and Future Work}

There are many open questions left to investigate.
One direction is to investigate deeper the different exegesis of nondeterminism, which would lead to a total combination of 16 possible exegeses, of which we conjecture that still only four of them reduce to two via Galois connections, thus ending up with 14 possible exegeses.
We would then like to investigate the impact of different assumptions about the program on this picture:
For example, what happens if we assume that executions started in $\pre$ always terminate?
What happens if all states in $\post$ are reachable?
What if the program is deterministic or reversible?

As for a different direction, none of the aforementioned triples expresses the existence of (at least) one path from~$\pre$ to~$\post$.
However, being able to more directly specify a bug like \enquote{it is possible to enter a wrong password and yet obtain successful login} would be quite desirable.
The following equations would express a property of this form:%
\begin{align*}
bpc \neq 0 \quad \text{and equivalently} \quad \wpp{\program}{\post} \land \pre \neq \text{false}
\end{align*}%
In contrast to the existing triples, this property speaks neither about \emph{all} initial states in $\pre$ nor \emph{all} final states in $\post$.

\bibliographystyle{ACM-Reference-Format}
\bibliography{refs}

\end{document}